\documentclass{article}

\usepackage{PRIMEarxiv}

\usepackage[utf8]{inputenc} 
\usepackage[T1]{fontenc}    
\usepackage{hyperref}       
\usepackage{url}            
\usepackage{booktabs}       
\usepackage{amsfonts}       
\usepackage{nicefrac}       
\usepackage{microtype}      
\usepackage{lipsum}
\usepackage{fancyhdr}       
\usepackage{graphicx}       
\usepackage{amsmath}
\graphicspath{{media/}}     

\title{Stochastic Model for Transfer of Gaseous Particles in Polymer/CNT Nanocomposites with Interfacial Regions 
}

\author{
  P.A.~Kalashnikova$^1$, I.Yu.~Kalashnikov$^2$, K.Yu.~Khromov$^1$ \\
  $^1$National Research Center ``Kurchatov Institute'', Akademika Kurchatova sq.1, Moscow, 123182, Russia\\
  $^2$Keldysh Institute of Applied Mathematics, 4 Miusskaya sq., Moscow, 125047, Russia\\
}

\begin{document}
\maketitle

\begin{abstract}
In this work a stochastic model of gaseous transfer in polymer/CNT nanocomposites is presented. The model takes into account interfacial areas, i.e.\@ polymer depletion regions. The local regime of transport is controlled by the density of the polymer. In a dense polymer, this regime corresponds to the ordinary diffusion, while in free volume regions it corresponds to the ballistic transport. 
The introduction of a free volume and/or a depleted polymer layer near to a CNT wall, leads to the emergence of anomalous diffusion.
We have demonstrated how the anomalous diffusion regime changes in the presence of nanotubes for different distributions of polymer density.
The presented approach allows us to describe the threshold effect in the diffusion coefficient as a function of CNTs density in polymer/CNT nanocomposites.
\end{abstract}


\section{Introduction}\label{sec:intro}

The study of particle transfer processes in heterogeneous systems of complex geometry is an importance problem in various fields such as nanoelectronics, chemical and membrane technologies, geology, aerospace, and medical fields. Examples of such systems are heterogeneous geological formations, oxide films growing on metal surfaces in aggressive media, cell membranes, and various composite materials, including polymer-based composites.

Polymers are widely used promising materials. They are robust, light, and resistant to various chemical phenomena, which led to their ubiquitous use today. However, as has been repeatedly demonstrated experimentally, the properties of polymers (including transport properties) can be significantly improved. This can be achieved by adding carbon nanofillers to the polymer, for example, carbon nanotubes (CNTs).

Transfer processes in polymer/CNT systems are directly related to their structure. 
As experimental and theoretical studies show, changes in the geometry of such systems are often expressed in a nonlinear increase/decrease in the coefficients of electrical conductivity, diffusion, permeability, etc \cite{Kim2, Grekhov, Holt, Bao, QZhang, Eremin1, Eremin2, Kim1, Khan, Likhomanova5, Sibatov}. Therefore, this leads to an efficient acceleration/deceleration of the transfer process. Obviously, it is extremely important for the manufacturing processes.

One of the practically important areas, where polymer/CNTs are applied is the field of membrane technologies. Here, so-called mixed matrix membranes, designed for the separation and purification of light gases, are used. In this work, we limited ourselves to considering specific systems -- PVTMS (polyvinyl trimethyl silane)/CNTs experiments \cite{Grekhov},\cite{Eremin2}. They studied permeability as a function of  CNTs fraction in the system and suggested the presence of a highly permeable layer of the modified polymer near the CNT surface with a characteristic size on the order of the diameter of nanotubes. In \cite{Eremin2} the thickness of membranes is $\sim 25~\mu$m, average CNT length is about $2~ \mu$m, and average outer CNT diameter is about $20-70$ nm. However, our model can be easily adapted to analogous systems -- for example, with polysulfone \cite{Kim2, Kim1} or polymethyl methacrylate (PMMA) \cite{Eremin1}.

The transport properties of such systems depend on many factors: the type of polymer, the density of the carbon fillers, the type of CNTs, and their distribution within the polymer. Dependence on a large number of parameters leads to the fact that the transport properties vary over a very wide range. Under these conditions, experimental studies aimed at finding nanocomposites with optimal properties can be very time-consuming and expensive. As a result, there is a need to develop an model for quantitative and qualitative prediction of the characteristics of nanocomposites.

In this article, we consider the process of diffusion transfer in polymeric CNT composites, namely, the transfer of light gases (H$_2$, N$_2$, CH$_4$, C$_3$H$_8$, CO$_2$, etc). It is known that there are three ways of transfer of gas particles: through the polymer matrix, through open CNTs (internal channel), and through interfacial regions resulting from poor adhesion between the polymer matrix and CNTs \cite{Grekhov, Kim2, Kusworo, Kim1}. In cases when the characteristic size of interfacial voids significantly exceeds the characteristic size of the CNT internal channels or when CNTs are closed, transport through CNTs can be neglected. Then interfacial regions are a key element in explaining the change in the mechanism of transport processes.

The incorporation of CNTs into a polymer, as noted above, is accompanied by an increase in the free volume due to poor adhesion between the polymer matrix and CNTs. This makes these volume elements easily permeable to gas particles. This assumption will be valid if the mean free path of a particle in a polymer is much smaller than the characteristic size of such a region \cite{Mao, Hongjun, Mantzalis}. Then the regime of particle transfer will be determined both by their collisions with the CNT walls, and by directional particle jumps between collisions, which should lead to a faster regime of particle transfer. Similar considerations regarding the change in the transport regime upon collision with CNT walls are given, for example, in \cite{Hongjun} and \cite{Bin-Hao} - in terms of Knudsen number. In \cite{Hongjun}, the effect of  superdiffusion of propane transport in open CNTs is demonstrated by the molecular dynamics method. As it was shown in \cite{Bin-Hao}, the hydrogen transfer regime depends on both the pore diameter and the particle loading and changes from single-file diffusion to ballistic.

The complexity of developing a gas transfer model is mainly due to the spatial heterogeneity of the system: under conditions of a strongly inhomogeneous medium, problems arise in the correct formulation and solution of transport equations.Methods such as \cite{Karger}, based on spatial homogenization, are associated with difficulties in numerically partitioning space. Also, various two-phase models, such as the Maxwell model \cite{Maxwell}, do not allow us to describe the non-linear threshold effect mentioned above. In some other works, for example \cite{Mao, Hongjun, Mantzalis, Bin-Hao, Mantzalis2, Ghoufi}, they try to model the transfer processes by the methods of molecular dynamics. However, all these works make it possible to reveal only local effects affecting transport but do not allow one to describe transport in micron-sized systems as a whole.

Existing theoretical models for various heterogeneous media \cite{Cherstvy, Metzler, Ghosh, Cherstvy2, Wang, Miyaguchi, Akimoto} (as far as the authors know, there are no such models for polymer/CNT systems), based on the use of the apparatus of random processes, almost always include some a priori assumptions associated with the use of the apparatus of non-Gaussian statistics. These assumptions are about a stable type of distribution of particles over jumps and/or waiting times between them, associated with the further use of the fractional differential apparatus. Such a priori assumptions have no physical justification, at least for our chosen system. Therefore, issues related to the appearance of stable distributions, namely, the physical mechanism leading to this type of distribution, in the polymer/CNT systems require a separate study. 

In the present study, we have developed a mathematical model with a complex particle transfer mechanism, which is a further development of the experience of the previous works \cite{Likhomanova2, Likhomanova1, Likhomanova4}. 
This mechanism combines the regimes of ballistic transfer, normal diffusion, and multiple collisions of a particle with CNT walls, depending on the localization of a particle in one or another region of the heterogeneous medium.
The proposed model introduces a spatial function that corresponds to the change in polymer density from the CNT wall to the polymer region (see \ref{sec:stochMod} for details). Then a system of stochastic equations is solved to find the increments of velocities and coordinates of a particle. Its solution corresponds to different transport regimes depending on the localization of the particle: ordinary diffusion in the polymer, reflection upon collision with the CNT wall and directional transfer near the CNT wall. The latter smoothly turns into ordinary diffusion as the particle moves away from a nanotube wall. Such a transfer mechanism seems to be physically justified based on the considerations outlined above.

\section{Stochastic Model}\label{sec:stochMod}

In our previous work on particle transport in polymer/CNT systems \cite{Likhomanova2}, we considered a one-dimensional formulation, where the following transfer mechanism was regarded. If the particle is in the polymer region, normal diffusion occurs, and in the regions corresponding to CNTs, ballistic transfer takes place. In this approach, it was impossible to take into account the reflection of particles from the tubes and related effects. Therefore, only the transfer of particles in the system, where the usual polymer alternates with free volume regions, was modeled. To simulate it, we considered the following system of stochastic differential equations (SDEs):

\begin{align}
	\begin{split}
		&dx={v}dt,  \\
		&d{v}=\chi(x)(-\gamma {v} dt +\sigma \delta {W}),
		\label{STCH1D}
	\end{split}
\end{align}

where $v$ – particle velocity, $\gamma$ – attenuation coefficient,
inversely proportional to the characteristic decay time, $W$ – Wiener process and $\sigma=const$ is the dispersion; the function $\chi$  takes the value of one for the polymer and smoothly but sharply goes to values close to zero for the free volume. 
A size of a transfered gaseous molecule is included into the parameter $\sigma$.
The function $\chi$ regulates the change of transfer regime.
At $\chi=0$, this is the ballistic transfer, $\chi=1$ corresponds to the usual diffusion. With this formulation of the problem, the anomalous diffusion was observed. 

The generalization of the system (\ref{STCH1D}) to the multidimensional case is quite obvious. However, to describe a real system, CNTs and elastic collisions of particles with their walls must be added to the model (see also~\cite{Hofling,Bodrova1,Bodrova2}). After a collision with a wall, a particle will acquire velocity $\mathbf{v}'=\mathbf{v}-2\mathbf{n}(\mathbf{v}\cdot\mathbf{n})$, where $\mathbf{v}$ is the velocity before the collision, {and} $\mathbf{n}$ is normal to a CNT wall. Since SDEs were supposed to be solved numerically, we have introduced the term $2\mathbf{n}(\mathbf{v}\cdot\mathbf{n})$ into the expression for the velocity increment with a certain coefficient $\mu_q$. This coefficient can be considered as an inverse time of the particle interaction with a wall. During the numerical calculations, $\mu_q$ was taken as an inverse time step, while the vector field $\mathbf{n}$ was set to be different from zero only in a small vicinity of the CNTs (see below). Thus, we wrote down a system of SDEs that is convenient for numerical calculations:

\begin{align}
\begin{split}
	&d\mathbf{r}=\mathbf{v}dt,  \\
	&d\mathbf{v}=\chi(\mathbf{r})(-\gamma \mathbf{v} dt +\sigma \delta \mathbf{W})-2\mathbf{n}(\mathbf{v}\cdot\mathbf{n})\mu_q dt,
	\label{STCH}
\end{split}
\end{align}

where $\chi(\mathbf{r})$ is
the function corresponding to the polymer density, and $\mathbf{W}$ are the independent Wiener processes.  
Here and below we consider a two-dimensional problem in Cartesian coordinates $(x,y)$.

\begin{figure}
\center{\includegraphics[width=1\linewidth]{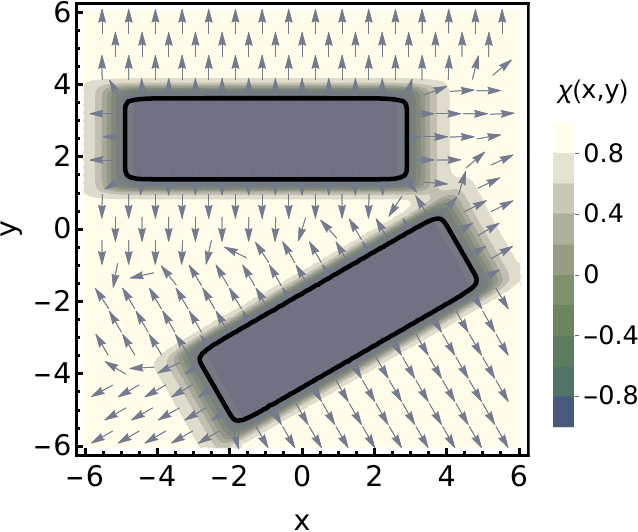}}
\caption{Schematic representation of CNTs, transition layer, and field of normals. The black contours correspond to the outer boundary of CNTs. The transition layer is shown with green shades. The normal vectors are shown by arrows.}
\label{fig:tubes}
\end{figure}

The SDEs (\ref{STCH}) describe the complex transfer mechanism of a particle in a heterogeneous polymer/CNT system, taking into account the reflection of a particle from the outer surface of a tube. The transition regions of polymer depletion are also considered: as a particle approaches a CNT wall from polymer, the rate of its velocity change decreases. Thus, the transfer regime is related to the definition of the spatial function $\chi$. In contrast to \cite{Mantzalis, Likhomanova1}, we do not assume any priority direction of particle motion near the tubes.


In this work, superellipses were chosen for the geometric representation of CNTs. The choice of the superellipses (instead of rectangles) is connected with the need of defining a smooth contour function. It allows us to define the normal vector field correctly.

Firstly, we defined the contour function $f(x,y,x_0,y_0,\theta_0)$ for any tube:

\begin{align}
\begin{split}
	f=\left( \frac{(x-x_0)\cos \theta_0 - (y-y_0)\sin \theta_0}{a} \right)^{2m} + \left( \frac{(x-x_0)\sin \theta_0 + (y-y_0)\cos \theta_0}{b} \right)^{2m} -1,
	\label{f}
\end{split}
\end{align}

where $(x_0, y_0)$ -- a geometrical center of a tube, $\theta_0$ -- a tilt angle, $a$ -- half-length of a tube and $b$ -- its half-width, $m$ is an integer number characterizing the roundness of the tube corners. The relation $f=1$ with fixed $x_0,y_0,\theta_0,m$ determines a superellipse. 

Next, we introduced an auxiliary function which is a smooth approximation of the Heaviside step function:

\begin{align}
\Xi= \frac{1}{2} - \frac{1}{\pi} \arctan\bigl(kf(x,y,x_0,y_0,\theta_0)\bigr),
\label{step}
\end{align}

where $k$ -- a coefficient that characterizes its sharpness. 
The function $\Xi$ takes the maximum value equal to one inside a tube and zero outside.

The $\chi$ function from (\ref{STCH}) should be defined in such a way that regions of polymer depletion appear near the outer surface of a nanotube.
The $\chi$ function should be designed so that $\chi=0$ at a CNT boundary and $0<\chi \leq 1$ outside a tube (see Fig.\ref{fig:tubes}). Inside a CNT $-1 \leq \chi<0$, however, this is not significant because due to the reflection from the walls, the particle cannot enter there. As mentioned earlier, we consider only the case of closed CNTs, so internal channels are not taken into account. Then $\chi$ may be determined by the product of the $\Xi$ functions for all tubes:
\begin{align}
\chi(x,y)=2 \cdot \prod_{i=1}^{N} \bigl( 1-\Xi_i(x,y,x_{0i},y_{0i},\theta_{0i},k_{\chi}) \bigr) -1,
\label{step}
\end{align}
where $N$ -- number of CNTs, $k_{\chi}$ is a sharpness parameter specific of $\chi$. In the limiting case of large values of $k_\chi$, $\chi$ is a step function and changes from $0$ to $1$ at a CNT boundary. Physically, this case corresponds to the absence of a free volume or to the strong adhesion of the polymer to a CNT. 
By varying the parameter $k_{\chi}$, one can introduce a transition region corresponding to a smooth change of the polymer density from the value equal $0$ to $1$ as one moves away from a CNT.
We can also shift the transition region in such a way that voids remain near the walls ($\chi = 0$), and at some distance from the wall, $\chi$ approaches to $1$. 

To solve SDEs (\ref{STCH}), it is necessary to define a normal vector field $\mathbf{n}$. For each tube, we defined a normal field as follows:

\begin{align}
\mathbf{n}_i=\frac{\nabla f_i}{|\nabla f_i|} \cdot \Xi_i(x,y,x_{0i},y_{0i},\theta_{0i},k_\text{norm}),
\label{normals}
\end{align}

where parameter $k_\text{norm}$ is responsible for the distance from the CNT boundary on which the normal field acts.  
To get the cumulative normal field $\mathbf{n}$, it is necessary to take the sum of $\mathbf{n}_i$ over all CNTs (see Fig.\ref{fig:tubes}). 
Thus, when a particle is away from a tube, the vector field $\mathbf{n}$ has no effect on it. When it approaches to a tube, the function $\chi$ becomes zero and only the second term in (\ref{STCH}) works, due to which the reflection occurs, because in the numerical calculations we chose values of the coefficient $\mu$ to be equal to the inverse time step.

Given the details described above, we solved (\ref{STCH}) using the Monte Carlo method. In the presented approach, a particle moves in a heterogeneous system: CNT/transition region/polymer and its mechanism of motion changes in accordance with its spatial position, as it is shown above.

\section{Numerical Modeling}\label{sec:NumMod}

To solve SDEs (\ref{STCH}), we used dimensionless quantities $l$ and $\tau$ which are the
characteristic parameters of the length and time, respectively. Dimensionless space parameters were chosen in accordance with the work \cite{Eremin2} where CNTs length is about $2 \cdot 10^{-4}$ cm, outer CNTs diameter is about $(2-7) \cdot 10^{-6}$ cm, and membranes thickness is about $(2.5-3) \cdot 10^{-3}$ cm and the typical diffusion coefficient in PVTMS (polyvinyltrimethyl silane) polymer is about $D=10^{-7}$ cm$^2$/s \cite{beckman}. Thus, we have chosen parameters $l=2 \cdot 10^{-4}$ cm and $\tau=1.25 \cdot 10^{-2}$ s. Further, we work in dimensionless quantities. In SDEs (\ref{STCH}) we took $\gamma=10$ and $\sigma=2.5$ to satisfy the relations for the time scale $\tau$ the diffusion coefficient
$ D = {\sigma^2}/{2\gamma^2}$.

Calculations were performed on a square system with the linear dimension $L=15$ with periodic boundary conditions. Firstly, the system is randomly filled with superellipses uniformly distributed in 2D space. All superellipses have an impermeable solid core with dimensions $a=0.5, b=0.025$ and parameter $m=7$ (see \ref{f}).
The CNTs density may be defined as $\eta=S_{se} \cdot N/L^2$, where $L$ -- the size of system, $N$ is the number of objects and $S_{se}$ is area of superellipse, which can be expressed via the gamma function $\Gamma$:

\begin{align}
S_{se}=4ab \frac{\Gamma{(1+1/m)}^2}{\Gamma{(1+2/m)}},
\label{se_area}
\end{align}

For each density we have generated $100$ random input configurations of nanotubes, where superellipses density was varied in the range of $0-0.324$. 
For a given configuration, the following trajectory implementation scheme was used.  
Initially, a particle is located at $(L/2,L/2)$. If this point is inside a CNT, then the algorithm searches for a point near the starting point that does not belong to the CNT. Then the particle moves according to the system of equations (\ref{STCH}) until the time $t=10^4$ is reached. 
In total, we performed $5 \cdot 10^5$ simulations: $5 \cdot 10^3$ trajectories for each of 100 spatial configurations of the superellipses.

\begin{figure}[]
\center{\includegraphics[width=1\linewidth]{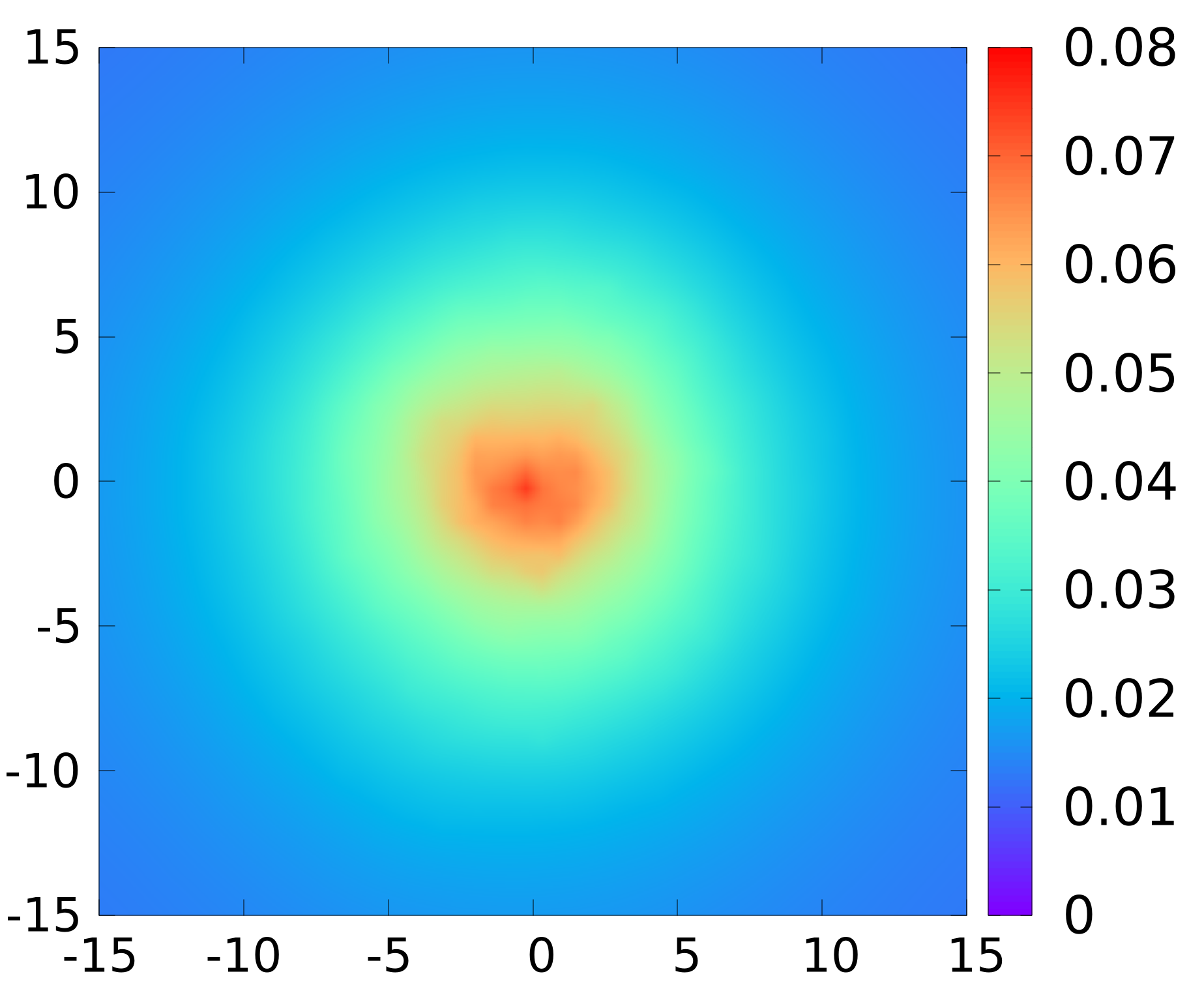}}
\caption{Two-dimensional distribution of the final positions of a particle at $t=10^4$ for $\eta=0.18$. }
\label{fig:distr3}
\end{figure}

At the output we get an offset vector from the starting point $(x-x_{start},y-y_{start})$, by which we build a two-dimensional distribution. The random process described by the system of equations (\ref{STCH}) must be axisymmetric on a large statistics (see Fig. \ref{fig:distr3}). Therefore we made a slice at $x=0$ and analysed it. 

The obtained distribution was fitted as a stable one, which has the following characteristic function \cite{nolan2016}:
\begin{equation}
\hspace{-0.15cm}\varphi(s)=\exp \left[i\delta s-\vert cs\vert^{\alpha} \left(1-i\beta\,  \text{sgn}(s) \tan\left( \frac{\pi \alpha}{2}\right) \right) \right],
\end{equation} where $-\infty<\delta<\infty$, $c\geq0$, $0<\alpha\leq2$ and $-1\leq\beta\leq1$ are the distribution parameters.
The characteristic exponent $\alpha$ determines the form of the distribution function, location parameter $\delta$ is the position of the maximum of the distribution, scale parameter $c$ is the width of the distribution, and the skewness parameter $\beta$ determines the distribution asymmetry.
Further, we analyzed the distribution fitting parameters for different types of the transition functions $\chi$ to estimate the influence of the local regime in near-wall areas and their size on the overall transfer regime.

\section{Results and discussion}\label{sec:Resu}

Since we consider a symmetric random process, the fitting of the obtained distributions shows that both parameters of skewness $\beta$ and location $\delta$ are about zero. 
The scale parameter $c$ does not influence on the form of a distribution. Therefore the most interesting parameter is $\alpha$, which is responsible for a difference of transfer regime from the ordinary diffusion.
Thus, to investigate the influence of the polymer depleted regions on the transfer regime, we have calculated $\alpha$ as a function of the CNTs density $\eta$ for three different functions $\chi$. 

\begin{figure}
\center{\includegraphics[width=1\linewidth]{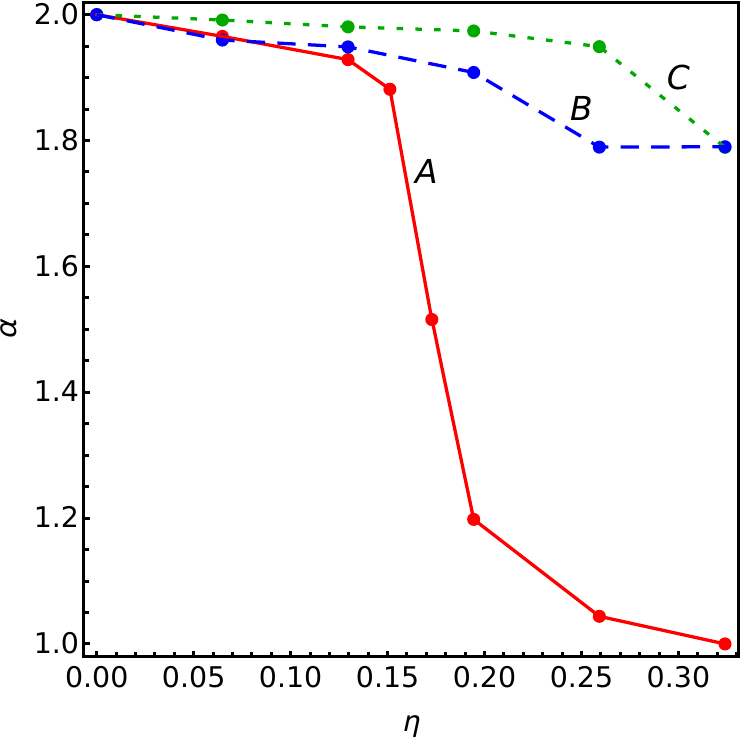}}
\caption{Stable distribution parameter $\alpha$ as a function of CNTs density $\eta$ for cases A (solid line), B (dashed) and C (dotted). }
\label{fig:percol}
\end{figure}

In the first case (case A), we considered a sharp form of $\chi$ with a shift. It was defined to be zero on the distance about three CNT diameters from a tube wall. Then it becomes unity, which corresponds to a huge value of $k_{\chi}$. In this case, for high densities $\eta$ almost all the space which is not occupied by CNTs has $\chi=0$. Hence, the transfer regime is mostly determined by collisions with the walls and ballistics. It is similar to a billiard ball moving in the geometry of random corridors. Then a particle should fly through these free volume corridors for distances much greater than those it would pass in the ordinary diffusion.
This leads to the strongly pronounced anomalous diffusion and a decreasing $\alpha$ with $\eta$ increasing, which is a consequence of $\langle r^2\rangle$ growth. At low densities, transfer regime is limited to the normal diffusion due to the lack of free volume changing the local regime. This can be clearly seen in Fig.~\ref{fig:percol}, where these results are presented.

In the case B we decreased the size of the near-wall area, setting $\chi$ to be zero at one CNT diameter. As can be seen from Fig.~\ref{fig:percol}, $\alpha$ decreases with increasing CNTs density $\eta$, reaching a value about $1.8$. However, this decreasing quite lower than in the case A. It is related to the fact that the particle spends more time in the polymer, where it moves according to the normal diffusion. 

In contrast to the first two cases, in the case C, a smooth transition from diffusion to ballistics regimes (smooth $\chi$) was considered. 
Here the free volume areas are absent, instead of them there are sparse regions whose density smoothly changes (see Eq.~\ref{step}). 
It reduces the effect of anomalous transfer for fixed density of tubes compared to the above cases.

We deliberately defined the density of the tubes in terms of "hard cores" without a permeable shell, as it is done in many percolation problems. It allowed us to estimate the effect of the size of the permeable region on transfer for identical structures. 
It is important to note that, as can be seen from Fig.~\ref{fig:percol}, the inflection point shifts to the right for each subsequent case (A, B, C).
Apparently, the largest drop of $\alpha$ is related to the percolation effect. A percolation cluster of free volume creates perforating ways for the particle transfer. According to percolation theory, this leads to the threshold effect. It cannot be described with a simple linear dependence (as in many two-phase models like \cite{Maxwell}) $D(\phi)=\phi \cdot D_1+(1-\phi) \cdot D_2$, where $D_1$ -- diffusion coefficient in polymer and $D_2$ -- diffusion coefficient in free volume.

The Einstein's relation is usually used to define the diffusion coefficient from experiments, $\langle r^2 \rangle \sim 2n D_{\text{eff}}t$ where $n$ is the space dimension ($n=2$ in our case). However, for anomalous transfer, diffusion coefficient must be defined in another way. For example, in \cite{zeleniy} it is suggested to use $\langle r^2 \rangle \sim 4 D \tau \left( {t}/{\tau} \right)^{\mu}$, where $\mu=2 \beta_s / \alpha$ and $\beta_s=1$ is a parameter responsible for the effects of memory. If we take $t=125$ s (a typical time of an experiment \cite{beckman}) and $\tau=1.25 \cdot 10^{-2}$ s (see Section~\ref{sec:NumMod}), then $D_{\text{eff}}/D \sim \left( {t}/{\tau} \right)^{\mu-1}$. For cases A and B $\alpha$ changes from $2$ to the minimum values $\alpha=1$ and $\alpha=1.8$ respectively. It corresponds to the change of $D_{\text{eff}}/D$ within all range of CNTs density $\eta$ in $10^4$ and $2.8$ times respectively. As far as the case A is concerned, such a low value of $\alpha=1$ is barely achievable in real experiments.
While the second value $2.8$ from the case B well agrees with experimental data for the oxygen molecules \cite{Eremin2}.
Moreover, this ratio is controlled by the parameter $\sigma$ and can be changed for different molecules sizes (see \ref{sec:stochMod}). Thus, we can carry out similar calculations for other gas molecules. 

This result is quite expected. Of all the cases we considered, it is case B, in which agreement with experiment was obtained, that best reflects the structure of the sample \cite{Eremin2}: when the size of the highly permeable channels is approximately equal to the diameter of the CNT (see Sec.~\ref{sec:intro}).
Thus, the presented 2D model allows us not only to qualitatively describe the nonlinear threshold effect in the diffusion (permeability) coefficient but also to approach quantitative estimates, taking into account transition layer at the CNT interface.

More accurate quantitative results may be obtained by considering a 3D problem and choosing $\sigma$ appropriately. Also, the $\mu$ exponent in 3D should differ from a 2D case. According to our preliminary estimates, when moving from 2D to a 3D model, the resulting effect should become less pronounced. However, this estimation must be tested in detail.

However, even using the two-dimensional model, we can claim a rough quantitative assessment of the transport characteristics. The use of the two-dimensional model in this work is associated with two aspects. Firstly, in the work \cite{Eremin2}, sizes of the composite sample were the following: the film thickness is about 25 $\mu$m, and the area is about 3.5 cm$^2$, i.e.\@ a thin film is used, which in a rough approximation can be considered as a 2D system. Secondly, numerical calculations of a similar three-dimensional model are possible at this stage of the implementation of our model, however they will still be quite expensive.

\section{Conclusion}

We have investigated the diffusion of gases in a heterogeneous environment, which reflects the transport properties of a polymer/CNT nanocomposites. Transfer regime of a particle is related to the complex geometry of the system and controlled by the CNTs density $\eta$ and a spatial function $\chi$. This function determines the density of the polymer and makes it possible to introduce a transition layer at the CNT interface and, therefore, take into account free volume regions (see Fig.~\ref{fig:tubes}). Our model is constructed to change the particle transfer regime when value of $\chi$ changes. The limiting cases of the particle transfer are the ordinary diffusion in the polymer and the ballistic regime in the free volume regions.

The influence of the three different shapes of function $\chi$ was analyzed. The results show the nonlinear dependence of the parameter $\alpha$ on CNTs density, which characterizes the deviation of transfer regime from ordinary diffusion. The change of this regime is most pronounced for case A, where the free volume regions have a size about three CNT diameters. For case B, when the size of free volume area is about one CNT diameter, this effect of transfer change is less than in the case A. 

Free volume regions were not considered in the case C. Here, the near-wall regions are represented by a soft transition of polymer density at the CNT boundaries, due to which $\alpha$ drop occurs at higher CNT densities (see Fig.~\ref{fig:percol}). Thus, in each of the three cases, anomalous diffusion is observed with an increase in the CNTs density.

The presented approach allows us to explain the threshold effect in diffusion coefficient as a function of CNTs density (see bottom of Sect.~\ref{sec:Resu}) and agrees well with experimental data \cite{Eremin2} for transport of oxygen in PVTMS/CNTs. 
This model can be easily extended in a number of directions. 
Firstly, it can be expanded to the three-dimensional case. We believe
that the appropriate 3D model will allow us to
obtain more accurate quantitative results for gaseous mass transfer in polymer/CNT nanocomposites. These results may be compared directly
to experimental data. Also, agglomeration effects \cite{Zhokh} can be included in our model. It is quite important aspect for the real experiments because usually they cannot be avoided fully.
Moreover, presented model may be applied to the other polymer types with incorporated CNTs, where the threshold effect for diffusion coefficient/permeability takes place \cite{Eremin1,Eremin3,Kim1}.

\section*{ACKNOWLEDGMENT}
The authors are grateful to K.V.~Chukbar, L.V.~Mateveev, R.T.~Sibatov for helpful discussions of the work.
This work has been carried out using computing resources of the federal collective usage center Complex for Simulation and Data Processing for Mega-science Facilities
at NRC “Kurchatov Institute” \cite{nrcki}.

\bibliographystyle{unsrt}  
\bibliography{references}

\end{document}